\documentclass[usenatbib]{mn2e}

\newcommand{\apj}{ApJ}           
\newcommand{\apjl}{ApJ}           
\newcommand{\mnras}{MNRAS}       
\newcommand{\nat}{Nature}
\newcommand{\aap}{A\&A}

\newcommand{\araa}{ARA\&A}
\newcommand{\aj}{AJ}
\newcommand{\pasp}{PASP}
\newcommand{\apjs}{ApJS}           

\usepackage{graphicx}
\usepackage{amssymb}
\usepackage{xcolor}
\usepackage{caption}
\usepackage{epstopdf}
\usepackage[flushleft]{threeparttable}
\usepackage{setspace}
\usepackage{times}
\usepackage[a4paper, total={17.8cm,24.0cm}, centering]{geometry}

\newcommand{\msun}{\hbox{$M_\odot$}}
\newcommand{\lsun}{\hbox{$L_\odot$}}
\newcommand{\atl}{ATLAS$^{\rm 3D}$}
\newcommand{\kms}{\hbox{km s$^{-1}$}}
\newcommand{\re}{\hbox{$R_{\rm e}$}}

\title[Galaxies star formation history at $z\approx0.8$]{Observed trend in the star formation history and the dark matter fraction of galaxies at redshift $z\approx0.8$}
\author[S. Shetty and M. Cappellari]{Shravan Shetty$^{1}$\thanks{E-mail:shravan.shetty@astro.ox.ac.uk} 
and Michele Cappellari$^{1}$\\
$^{1}$Sub-Department of Astrophysics, Department of Physics, University of Oxford, Denys Wilkinson Building, Keble Road, Oxford OX1 3RH, UK}

\begin{document}
\singlespace

\pagerange{\pageref{firstpage}--\pageref{lastpage}} \pubyear{2015}

\maketitle

\label{firstpage}

\begin{abstract}

We study the star formation history for a sample of 154 galaxies with stellar mass $10^{10}\lesssim M_{\ast} \lesssim 10^{12} \msun$ in the redshift range $0.7 < z < 0.9$. We do this using stellar population models combined with full-spectrum fitting of good quality spectra and high resolution photometry. For a subset of 68 galaxies ($M_{\ast}\gtrsim 10^{11}\msun$) we additionally construct dynamical models. These use an axisymmetric solution to the Jeans equations, which allows for velocity anisotropy, and adopts results from abundance matching techniques to account for the dark matter content.
We find that: (i) The trends in star formation history observed in the local universe are already in place by $z\sim1$: the most massive galaxies are already passive, while lower mass ones have a more extended star formation histories, and the lowest mass galaxies are actively forming stars; (ii) we place an upper limit of a factor 1.5 to the size growth of the massive galaxy population; (iii) we present strong evidence for low dark matter fractions within 1\re\ (median of 9 per cent and 90th percentile of 21 per cent) for galaxies with $M_{\ast} \gtrsim 10^{11} \msun$ at these redshifts; and (iv) we confirm that these galaxies have, on average, a Salpeter normalisation of the stellar initial mass function.

\end{abstract}


\section{Introduction}

Currently, one of the key quests of Astrophysics is to understand and model the processes that guide the formation and evolution of galaxies. Great strides have been made over the past few decades and with the advancement of technology, such as ever larger telescopes taking ever larger surveys of thousands of galaxies within an ever larger redshift range \citep[e.g.][]{SDSS,2dFGRS,VVDS,DEEP2,PRIMUS,AGES}, the advent of new techniques such as gravitational lensing \citep[e.g.][]{SLACS,CFHTLenS}, and galaxy surveys using integral field spectroscopy \citep[e.g.][]{SAURONsurvey,atlas3d1,CALIFA,Scottetal2015,MANGA}.

Recent observational evidence suggests that the star formation rate of the universe peaked at $z\sim2.5$ and that by $z\sim1$ half of the stellar mass of the universe today was already in place \citep{BundyEllisConselice2005,Mortlocketal2011,Muzzinetal2013}. The decreasing star formation rate, referred to as quenching, is mass dependent with the more massive galaxies being quenched earlier. Also, the comparison of the most massive galaxies ($M_{\ast}\gtrsim10^{11}\msun$) at high and low redshifts show that these quiescent galaxies have undergone a size evolution; with the size of the galaxies increasing with decreasing redshift \citep{Daddietal2005,Trujilloetal2006,VanDerWeletal2008,Bezansonetal2009,vanDokkum2010z2}. This size evolution has been associated with minor mass growth, suggesting that these growths may be driven by minor merger where the size of the galaxy grows to the second power of the added mass through virial arguments, unlike major mergers where the size grows linearly to the increase in mass \citep{Naabetal2009,Bezansonetal2009,VanDerWeletal2009}. Additionally, recent works have pointed out that a significant part of the observed size growths in the populations of quiescent galaxies, especially at lower masses, may be due to progenitor bias, wherein the addition of large recently quenched galaxies contribute to the observed increase in the mean size of the population  \citep[e.g.][]{Carolloetal2013,Poggiantietal2013}. 

Regardless of what the process for the growth of the galaxy size, and its stellar mass may be, there is strong evidence indicating that, for the most massive galaxies, most of the additional stellar mass is added to the outskirts of the galaxies, while the central regions remain mostly unperturbed \citep{Trujilloetal2006,Bezansonetal2009,vanDokkum2010z2}. The end result of this merging process are the most massive galaxies in the nearby Universe which are found to be slowly rotating \citep{ATLAS3D3,atlas3d20}, they have cores in their surface brightness profiles \citep{Faberetal1997,Ferrareseetal2006,Atlas3D23}, and are embedded in extended stellar envelopes \citep{Kormendyetal2009}.

The situation appears radically different for less massive ($M_{\ast}\la10^{11}\msun$) passive galaxies. At the present day, they are structurally different, and appear to have followed a different evolution path \citep{atlas3d20}. They are axisymmetric \citep{Atlas3D_2}, they contain disks like spiral galaxies \citep{Atlas3D_7} and are dominated by rotation \citep{ATLAS3D3,atlas3d20}. These fast rotating galaxies follow the same mass-size relation, and have the same mass distribution, both in dense clusters as in the field  \citep{Cappellari2013ApJL}, indicating they experienced an insignificant amount of merging during their evolution, in agreement with redshift evolution studies \citep{vanDokkumetal2013}.

Due to the recent advances in the techniques of stellar population modelling and redshift surveys, a key addition to this emerging picture of galaxy evolution is provided by studies of the stellar populations of galaxies through cosmic time. 

The work of \citet{schiavon2006}, using spectra from the DEEP2 survey \citep{DEEP2}, compared to local SDSS \citep{SDSS} results, suggests that the evolution of the red-sequence galaxy population is not consistent with a passive evolutionary model. Instead, they propose that the red-sequence population should either continue to host some level of star formation (``Frosting'') to present day or have newly quenched galaxies joining the red-sequence galaxies between $z\sim0.9$ and today.

\citet{Choietal2014} study quiescent high redshift galaxies via a full spectrum fitting of stacked galaxy spectra to derive the stellar ages and elemental abundances of Fe, Mg, C, N and Ca. The work uses optical spectra of local galaxies taken from the SDSS and spectra from the AGES \citep{AGES} survey within a redshift range of $0.1<z<0.7$. They find negligible evolution in elemental abundances at fixed stellar mass. For the most massive galaxies they measure an increase in stellar age consistent with passive evolution since $z\approx0.7$. While at masses below $\sim10^{10.5}$ \msun, the data permit the addition of newly quenched galaxies.

\citet[Hereafter G14]{Gallazzietal2014} study a sample of 70 quiescent and star-forming galaxies at $z\approx0.7$, above a stellar mass of $10^{10}\msun$. They derive the stellar age-mass relation of the galaxies, which they compare with the one derived in a similar manner in the local universe. They find that taken as a whole, passive evolution cannot represent the evolution of galaxies in the last $\sim6$ Gyr. In fact, although the shape of the stellar age-mass relationship between the two redshifts is similar, the offset is inconsistent with passive evolution. This is agreement with their observed metallicity differences with redshift. They propose a mass-dependent star formation history (SFH) to explain the observations. 

Here we use full-spectrum fitting to explicitly determine trends in the star formation history of a sample of 154 galaxies at $0.7<z<0.9$. Furthermore, we investigate the correlation between the stellar population and the physical parameters of the galaxies. We also present results on the dynamical modelling of a subset of 68 galaxies. This subsample is the same we analysed in our previous work \citep{ShettyCappellari2014ApJL}, where we studied the Initial Mass Function (IMF) mass normalisation and concluded it is consistent with a \citet{salpeter1955} slope. Here, we improve upon the dynamical models by accounting for the dark matter of the galaxies via abundance matching techniques. 

In Section~2 of the paper, we describe the observational data that we use within this study while in Section~3 we discuss the selection criteria that have been implemented in the course of this analysis, along with a comparison of our galaxy sample with the parent sample. In Section~4, we present the various methods used to analyse the data followed by a discussion of the results in Section~5. In Section~6, we provide a summary of the results.

In this study, we assume a flat universe with the following cosmological parameters: $\Omega_{\rm m}=0.3$, $\Omega_{\rm \Lambda}=0.7$, $H_{\rm 0}=70$ \kms\ Mpc$^{-1}$.

\begin{figure*}
	\centering
	\includegraphics[width=0.8\linewidth]{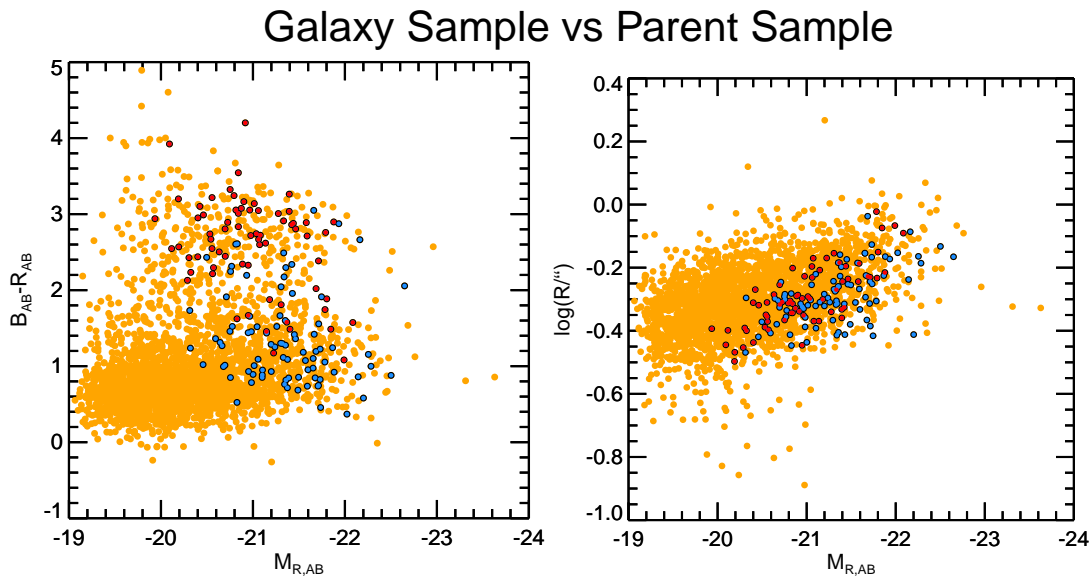}
	\caption{In this plot, we compare our galaxy samples: 68 galaxies in our secondary sample (red points) and 154 galaxies in our full sample (blue and red points), against the parent sample of $2,896$ galaxies in the same redshift range and field from the DEEP2 survey (orange points). In the left panel, we present the colour-magnitude plot where we can distinguish between the red sequence and blue cloud of the parent sample. It is evident that our galaxy samples have a measurably higher magnitude cut-off, compared to the parent sample, which is due to our selection criteria. Within this higher magnitude cut-off, our full galaxy sample provides a good representation of both the red sequence and the blue cloud. We also notice that our secondary galaxy sample is a good representation of the red sequence, while also including a few very luminous galaxies of the blue cloud. If we assume that the $R_{\rm AB}$ magnitude is a tracer of stellar mass, the right panel can be approximated as a stellar mass-size plot for the galaxies. We see here that our galaxy samples mainly consist of galaxies above a certain surface brightness.}
	\label{Sample_plot}
\end{figure*}

\section{Data}

\subsection{Spectral Data}

The spectral data used in this study was taken from the DEEP2 survey. The survey is an apparent magnitude limited, $R_{AB}=24.1$, spectroscopic redshift survey of 4 fields covering an area of $2.8$ deg$^{2}$ across the sky characterised by low extinction. The observation targets for the survey were selected using an algorithm based on the Canada-France-Hawaii Telescope (CFHT) observed object size, luminosity, (B-R) and (R-I) colours \citep{Coil2004} and a surface brightness cut-off which eliminated wrongly classified objects \citep{DEEP2}. Of the four fields of DEEP2, only the Extended Groth Strip (EGS; \citet{groth1994}) has the required high resolution {\it HST} data to construct dynamical models and accurately measure galaxy siz. Hence this study is restricted to this field of the DEEP2.

The observations for the survey were done using the DEIMOS spectrograph on the Keck-2 telescope, covering an observed wavelength range of 6,100\AA ~- 9,100\AA. The highest resolution grating, 1,200 line mm\textsuperscript{-1}, was used to achieve a spectral resolution of R $\sim6,000$ within the $1\arcsec\times\sim7\arcsec$ slitlets of the instrument. The observations were taken in 3 exposures of 20 minutes each, with the FWHM generally ranging between $0\arcsec .5$ and $1\arcsec .2$ due to seeing.

The reduction of the 2D spectrum of each galaxy and the extraction of its 1D spectra was done by the DEIMOS {\em spec2d} pipeline. In our study we use the 1D spectra extracted via the Boxcar technique from the central $\sim 1\arcsec \times 1\arcsec$ window. Due to the design of the survey, the observed 1D spectra is divided into 2 halves along wavelength. Here we use the blue half of the two, since the CaK (3,934\AA) and G band (4,304\AA) stellar lines fall within this half for galaxies in the redshift range that we are studying.

\subsection{Photometric Data}

Our photometric data was taken from the AEGIS data product \citep{aegis}, specifically from the {\em Hubble Space Telescope} ({\it HST}) GO program 10134 (P.I.: M. Davis). These are {\it HST}/ACS images taken in the F606W (V band) and F814W (I band) filters. The observations are limited to 5$\sigma$ measurements of $V_{\rm F606W}=26.23(AB)$ and $I_{\rm F814W}=25.61(AB)$ within a circular aperture of $0\farcs3$. This program covered $0.197$ deg$^{2}$ of the EGS, overlapping entirely with the region observed by DEEP2. The observations were made in a four point dither pattern, which was processed by the STSDAS MultiDrizzle package to produce a final mosaic of the field with a pixel scale of $0\farcs03$. In this work, we use the F814W photometry as this translates to the vega B band at the redshifts that we are studying.

\section{Galaxy Samples}

In this study we use two samples extracted from the large DEEP2 survey catalog. The first one consists of 154 galaxies and was used to measure trends in the galaxies star formation histories via full-spectrum fitting. The second one is a subsample of 68 galaxies of the first, for which we construct dynamical models. Details of the adopted selection criteria are presented below.

\label{sample}

\subsection{Selection Criteria}
\label{SelectionCriteria}

Our first selection criterion is based on the quality flags provided by the DEEP2 survey catalogue, i.e. we selected galaxies that have either ``secure'' or ``very secure'' redshift estimation based on the automated and visual spectral redshift estimation by the DEEP2 team. We select galaxies with redshift $0.7<z<0.9$. In this study we have only used galaxies in the EGS field due to the lack of availability of high resolution {\it HST} photometry in the remaining fields. By extracting $9\arcsec \times 9\arcsec$ thumbnails, from the AEGIS mosaics, centred at the galaxy coordinates provided by the DEEP2 catalogue, we visually confirm the presence of photometry of the galaxies. These thumbnails are later used to parametrize the photometry of the galaxies and further details on this is provided in Section~\ref{QualMGE}. From a total of 43,569 galaxies in the DEEP2 Galaxy Redshift survey 1,249 galaxies passed this stage of the selection process. Of these, 1,240 ($99.3\%$) galaxies have an apparent $I_{\rm F814W} < 24(AB)$.

The second selection criterion is based on the quality of the galaxy spectra. We calculate the mean signal-to-noise ($S/N$) of a galaxy spectrum as the median flux in the galaxy spectrum divided by the standard deviation of the residuals of the spectral fit to that spectrum. In this instance, the spectral fitting was done with a set of empirical stellar spectral library, while masking gas emission features. Further details on this spectral fitting is presented in Section~\ref{QualSpec}. We only select galaxies with $S/N>3$. We further eliminate galaxies that do not appear to have significant absorption features, i.e. clear spectral absorptions which stood out visually above the noise. This selection criteria reduced the galaxy sample to 175 galaxies.

Our third selection criteria is based on the visual inspection of the thumbnails of the galaxies, taken from the F814W images by the {\it HST}/ACS. We have removed 21 irregular galaxies from our final sample as these galaxies are likely to be disturbed and would confuse the trends being investigated in this study. The selection criteria based on the galaxy spectra and imaging quality, dramatically reduces our galaxy sample to 154 galaxies.

In this study, we also create dynamical models for our galaxy sample, however this requires some additional selection criteria. To create the dynamical models we need to parametrize the photometry of our galaxies to high accuracy within the central regions from which the stellar kinematics are observed. We exclude galaxies that have disturbed photometry, such as dust lanes, or non-axisymmetric features, and hence this galaxy subsample is biased towards early-type galaxies with smooth light profiles. Further details on this selection criterion is given in Section~\ref{QualMGE}. 

Also, the dynamical modelling of galaxies assume a spatially constant stellar mass-to-light ratio ($M/L$), however this assumption becomes inaccurate for galaxies with multiple significant star formation events. Therefore, after deriving the SFH of our galaxies, we eliminate galaxies that require more than one significant star formation episode to reproduce their spectrum. In addition to this, the \atl\ group found that young galaxies, as inferred by their strong $H\beta$ absorption feature, tend to show strong gradients in the stellar $M/L$. For this reason, these galaxies where removed from their IMF studies \citep[hereafter C13b]{cappellari2012nature,atlas3d15}. In this study, we use full-spectrum fitting to the galaxy spectra to identify the best fitting single stellar population (assuming solar metallicity) of the galaxies. We then classify any galaxy with the best fitting population of 1.2 Gyrs or younger as a young galaxy and eliminate them from our secondary sample. This narrows our secondary galaxy sample to 68 galaxies for which dynamical models were created. This sample is identical to that used in \citet{ShettyCappellari2014ApJL}.

\subsection{Sample Bias}
\label{bias}

Given our strict selection criteria, our sample cannot be representative of the general galaxy population at $z\sim1$. To illustrate the effect of the selection criteria on the sample, in Fig.~\ref{Sample_plot} we compare our galaxy samples to a larger parent sample of galaxies. This parent sample contains $2,896$ galaxies selected to have ``{\it secure}'' or ``{\it very secure}'' redshift $0.7<z<0.9$ in the EGS field of the DEEP2 survey. We compare the galaxy samples using data from the photometric catalogue provided by the DEEP2 survey, which is derived from \citet{Coil2004}. The catalogue contains the observed $B_{\rm AB}$, $R_{\rm AB}$ and $I_{\rm AB}$ magnitudes of the galaxies calculated from observations made by Canada-France-Hawaii Telescope using the CFH12K camera. The catalogue also contains an estimate of the Gaussian radius of the galaxies which we use as a proxy for the size of the galaxies. For further details on the the catalogue, we refer the reader to \citet{Coil2004}.

In the left panel of Fig.~\ref{Sample_plot}, we compare the $B_{\rm AB}-R_{\rm AB}$ colour of the galaxies against their absolute $R_{\rm AB}$ observed band magnitude. The plot depicts the $R_{\rm AB}$ magnitude limit of the DEEP2 survey and clearly separates the blue cloud and red sequence of the field. The plot demonstrates that our sample have a higher $R_{\rm AB}$ cut-off than the DEEP2 which can be explained by the S/N cut-off that we place on our galaxy spectra. This is further demonstrated in the right panel of Fig.~\ref{Sample_plot}, where we plot the $R_{\rm AB}$ of the galaxies against a measure of the size of the galaxies. Due to the design of the survey, particularly the fact that each galaxy has a fixed exposure of $3\times 20$ minutes and fixed aperture size, galaxies with lower surface brightness will be of lower quality. Hence, our S/N cut-off can be roughly approximated as a surface brightness cut-off.

The left panel of Fig.~\ref{Sample_plot} also demonstrates that our secondary sample, indicated by the red points in the plot, is a good representation of the red sequence along with a few galaxies in the blue cloud. This is likely the consequence of the additional selection that we place on the secondary sample, whereby only galaxies with smooth light profiles and without significant star formation younger than 1.2 Gyrs are selected. Our full sample, indicated by the red and the blue points, appears to represent all but the bluest galaxies over a certain magnitude cut-off. This under-representation may be the result of the visual inspection of the galaxy spectra, where the presence of hydrogen emission or apparent high frequency noise make the spectrum fits look unreliable. 

\section{Methods}

\subsection{Spectrum Quality Criteria}
\label{QualSpec}

We measure the mean $S/N$ by fitting the observed galaxy spectra with an empirical stellar spectral library. Our choice for an empirical stellar library, rather than population models, to estimate the $S/N$ is done to allow for maximum freedom to the optimal template combination, especially for very young ages. This avoids potential bias in the stellar velocity dispersion, which is also extracted during this step. However our results are quite insensitive to this choice, due to the rather low $S/N$ of our spectra. The spectral fitting is done using the pPXF method and code \citep{ppxf} which uses a penalised maximum likelihood fitting technique in pixel space. Before fitting we logarithmically rebin the observed galaxy spectrum to 60 \kms\ per pixel. As templates we use a subset of empirical stellar spectra taken from the Indo-US Library of Coud\'{e} Feed Stellar Spectra Library \citep{indo-uslibrary}. The subset of 53 templates is made such that: (i) each spectrum is gap-free and (ii) the subset is a good representation of the library's atmospheric parameter range ($T_{\rm eff}$ vs $[Fe/H]$). We use pPXF to fit for the velocity and velocity dispersion of the galaxies using 4 additive and one multiplicative polynomial. During the fit of the galaxy spectra gas emission features are masked. The residuals of the fits are found to be random, suggesting that our spectral fitting is not biased by template mismatch.

Using these fits, we derive the $S/N$ of each galaxy spectra as the ratio of the median of the galaxy spectra and the standard deviations of the residuals. We remove galaxies that have $S/N<3$ from our final sample as we found the fits to their spectra unreliable.

\subsection{Recovering Star Formation History}
\label{sec_SFH}

We derive the star formation history of our main sample of 154 galaxies via a full-spectrum fitting technique using the pPXF code \citep{ppxf}. For our templates, we use the MILES stellar population models \citep{milesmodels}. These models are based on the MILES empirical stellar spectra library \citep{mileslibrary} and have a resolution of 2.5 \AA\ FWHM \citep{Falcon-Barrosoetal2011}. To ensure realistic fits of the galaxy spectrum with the template models, we placed certain constraints on our template grid. We use stellar population models logarithmically spaced into 40 age bins within the age range of 0.08--7.9 Gyrs to restrict the age of stellar populations to that less than the age of the universe at the redshift under study. We have also restricted our template metallicities to [M/H] of -0.4, 0.0 (Solar metallicity) and 0.22. This constrain is based on results by \citet{schiavon2006} where the authors studied the absorption line strengths of stacked spectra of the DEEP2 sample at redshifts $0.7<z<1$ and concluded that these galaxies have metallicities close to solar. Since \citet{schiavon2006} use [Fe/H] to measure their metallicities, we should note that for our template models the [M/H] is equivalent to [Fe/H] \citep{milesmodels}. For this analysis we adopt the Salpeter IMF \citep{salpeter1955} for our stellar population models, however the derived trends in the SFH are insensitive to the IMF choice, as demonstrated in \citet{Panteretal2007}.

In this study, we do a full-spectrum fitting for the spectral features and continuum shape of the observed galaxy spectra. This is done using four multiplicative polynomials to account for effects of reddening and calibration uncertainties in the observed galaxy spectra. The fractional weight assigned to each template indicates the fraction of the galaxy spectrum contributed by that stellar population. We do not use any additive polynomials when fitting for the stellar populations as these polynomials can artificially change the line strengths of the model templates. During the fitting with pPXF, the code convolves the model spectra with a Gaussian Line-Of-Sight Velocity Distribution (LOSVD), which is optimized to best fit the spectrum.

\begin{figure}
  \centering
	\includegraphics[clip, width=0.9\linewidth]{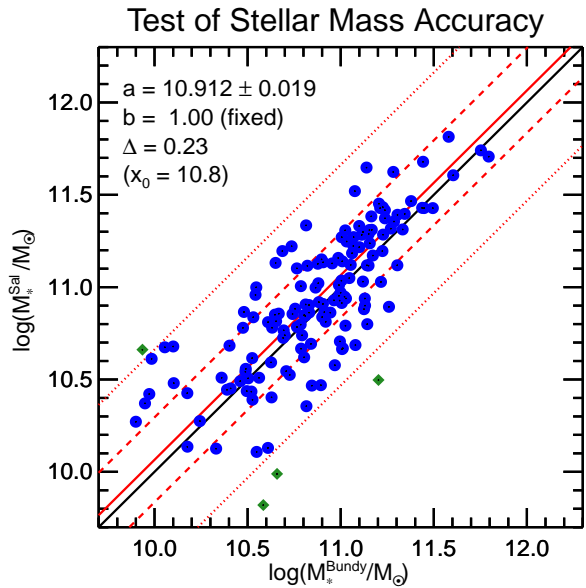}
	\caption{Here we present a comparison between the stellar mass derived via our population modelling with the same derived by \citet{bundy2006}. The linear fitting is done via Least Trimmed Square fitting described in C13b with the code modified to fit the slope at one. The best-fitting coefficients for the fitted relationship, $y=a + b(x-x_{0})$, are presented in the top left corner of the plot. The solid red line is the best fitting relation between the two quantities, while the dashed and dotted red lines enclose regions of $1\Delta$ and $2.6\Delta$ respectively, where $\Delta$ is the observed rms scatter around the relation. For comparison, the solid black line represents the one-to-one relation between the quantities.}
	\label{Bundy}
\end{figure}

We derive the star formation history of the galaxies by fitting their spectra with the above mentioned template grid while enforcing a smoothness criterion on the distribution of the weights in a mass weighted manner. This smoothness criterion is enforced by penalising the $\chi^{2}$ for weight distributions with non-zero second partial derivatives. In this way, for infinite regularization, the 2-dimensional solution converges to a plane. Regularization is a standard method to solve general ill-posed inverse problems \citep[e.g.][]{engl1996regularization} like the recovery of the SFH from a spectrum. It is implemented in pPXF using the classic linear-regularization approach as in equation~(19.5.10) of \citet{NumericalRecipes} and is enforced via the {\it REGUL} keyword in pPXF. The value assigned to the {\it REGUL} keyword specifies the strength of the penalisation of the $\chi^{2}$. A common guideline is to penalise the model till $\Delta\chi^2=\chi^{2} - \chi^{2}_{\rm unregul} = \sqrt{2\times N_{\rm DoF}}$, where $\chi^{2}_{\rm unregul}$ is the $\chi^{2}$ of the model without penalisation and $N_{\rm DoF}$ is the number of degrees of freedom \citep[section~19.5]{NumericalRecipes}.

We should note that the regularization approach is not similar to smoothing the solution after the fit. In fact regularization {\em does not} prevent sharp features in the recovered star formation or metallicity distribution, as long as these are required to fit the data. The regularization only influence the weight distribution when the solution is degenerate and many different and noisy ones can fit the data equally well. In this case the smoothest solution is preferred over the noisy ones. This is a general feature of all regularization approaches.

For a few galaxies, we find the value of {\it REGUL} is poorly constrained by the $\Delta\chi^{2}$ criterion, due to the rather large noise in the galaxy spectra. For this reason pPXF is able to fit the spectrum using a perfectly smooth weight distribution before $\Delta\chi^{2}= \sqrt{2\times N_{\rm DoF}}$. Hence, to be conservative, and to avoid our results from being driven by the regularization, we lowered the requirement on $\Delta\chi^{2}$ from the standard one mentioned above to $\Delta\chi^{2} = \sqrt{2\times N_{\rm DoF}}/10$.

As a consistency check, we have compared the stellar masses of 151 common galaxies derived in this study and that of \citet{bundy2006}, where the authors use BRIK colour to derive the stellar mass for a sample of $\sim8,000$ galaxies in the DEEP2 survey under the assumption of a Chabrier IMF \citep{chabrier2003}. We calculate our galaxy stellar mass ($M_{\star}^{Sal}$) by multiplying the absolute luminosity of the galaxies by the $(M_{\ast}/L)$ inferred from the spectral fits
\begin{equation}
	(M_{\ast}/L)_{\rm Sal}= \frac{\sum_{j}w_{j}M_{\star, j}^{\rm no gas}}{\sum_{j}w_{j}L_{{\rm B},j}},
	\label{MLeq}
\end{equation}
where $w_{j}$ is the weight attached to the $j$th template by the regularised mass-weighted fit to the galaxy spectrum, $M_{\star, j}^{\rm no gas}$ is the mass of the template that is in stars and stellar remnants, and $L_{{\rm B},j}$ is the $B$-band (Vega) luminosity of the template. Please note that all M/Ls derived in this study are in the B (Vega) band. We derive this luminosity using the MGE parametrization of the {\it HST}/ACS galaxy thumbnails, mentioned in Section~\ref{QualMGE}, using the galaxy redshift and equation~(10) of C13b.

The comparison between the 151 common galaxies is shown in Fig.~\ref{Bundy}, where we compare the two stellar masses using the \textsc{lts\_linefit} method \footnote{Available at http://purl.org/cappellari/software} (C13b) which has been modified to derive the best fit with a fixed slope. The solid red line in the plot represents the best fitting unitary-slope relation to the points, with the coefficients presented in the top left corner. The $\Delta$ parameter is the observed rms scatter around the relation, while the dashed and dotted red lines indicate the $\Delta$ and $2.6\Delta$ deviations from the best-fit. The solid black line indicates the one-to-one relation of the quantities. We see that a relation with slope unity is a good fit between the two datasets, though there is a significant offset of 0.11 dex and some scatter. The cause of the scatter and offset is likely the effect of differences in the assumed IMF between the two quantities, ie this study assumes a Salpeter IMF while \citet{bundy2006} assume a Chabrier IMF, and the differences between the photometry and its analysis. Also, the work by \citet{GallazziBell209} found that based on the techniques used to model the $(M_{\ast}/L)$ of the galaxies, uncertainties and systematics of the order 0.1--0.2 dex are expected.

Given that the stellar masses derived in this study and that of \citet{bundy2006} are completely independent, use very different data and approaches, and suffer from different systematics, we adopt $\Delta/\sqrt{2}=0.16$ dex as our very conservative errors in the estimation of the stellar mass of an individual galaxy via stellar population modelling.

\begin{figure}
	\centering
	\includegraphics[clip, width=0.9\linewidth]{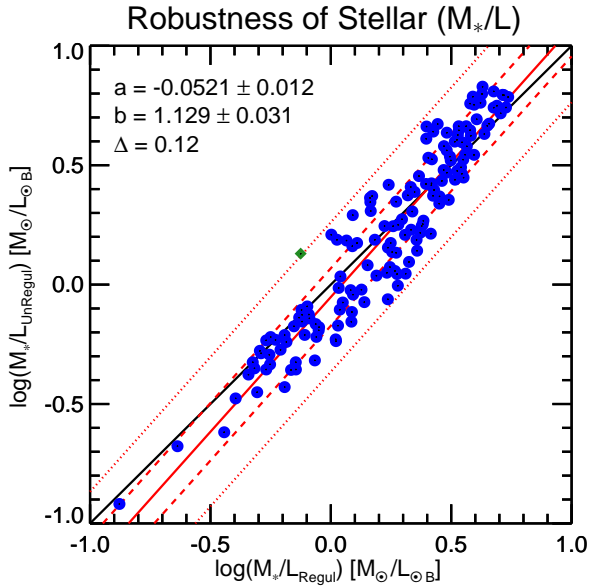}
	\caption{In this plot, we compare the stellar M/L derived under two different assumptions for the stellar population of the galaxies. The comparison between the derived quantities is done using the Least Trimmed Square (LTS) fitting described in C13b. The best-fitting coefficients for the fitted relationship, $y=a + bx$, are presented in the top left corner of the plot. The solid red line is the best fitting relation between the two quantities, while the dashed and dotted red lines enclose regions of $1\Delta$ and $2.6\Delta$ respectively, where $\Delta$ is the observed rms scatter around the relation. The solid black line is the one-to-one relation between the quantities. The independence of the quantities of systematic bias indicates that our fits do not suffer from template mismatch.}
	\label{regulvsunregul}
\end{figure}

To estimate a conservative uncertainty in our derived $(M_{\ast}/L)_{\rm Sal}$, we derive our $(M_{\ast}/L)$ under 2 model sets and fitting conditions: (i) a regularised fitting to the complete model grid of 120 single stellar population models and, (ii) a non-regularised best fit to the model grid limited to 40 models with solar metallicities. Our comparison of the two in Fig.~\ref{regulvsunregul} shows a linear relation between the two values. The observed scatter is $\Delta=0.12$ dex, which corresponds to an error of $\Delta/\sqrt{2}=0.08$ dex or 22\% in each $(M_{\ast}/L)$, assuming their errors are comparable.

For comparison we also derived the error on the derived $(M_{\ast}/L)_{\rm Sal}$ via a more standard bootstrapping technique. We reshuffled the residual and repeated the determination of $(M_{\ast}/L)_{\rm Sal}$ multiple times. This yielded a median error of 0.05 dex in $(M_{\ast}/L)_{\rm Sal}$ for the full sample, which is smaller than the error of 0.08 dex derived above. To be conservative, and to account at least in part for systematic effects, in this study we adopt the intrinsic scatter of 0.08 dex between the two quantities as the error in the derived $(M_{\ast}/L)$.

\subsection{Photometry Parametrization}
\label{QualMGE}

To study the scaling relations, we measure the luminosities and size by parametrising the photometry of the galaxies. In this study, we choose the F814W photometry taken by the {\it HST}/ACS to do so. The need for {\it HST} photometry comes from the requirement, in the dynamical models, of an accurate description of the tracer population producing the observed kinematics. This also motivates our choice for the F814W filter, which is as close as possible to the wavelength range from which the stellar kinematics is obtained.

We parametrize the photometry of the galaxies using the Multiple-Gaussian Expansion technique \citep{emsellem1994,cappellari2002}. This technique fits a series of two dimensional Gaussians, each defined by an amplitude, standard deviation and axial ratio, to the observed surface brightness of the galaxy image in a non-parametric manner, i.e. without the use of a predefined light profile function like the S\'{e}rsic profiles. The advantage of this technique is that the convolution and deprojection (i.e. conversion of the 2D model into the intrinsic 3D luminosity distribution) of the MGE models is analytically simple. The fit is performed with the robust method and software (see footnote 1) by \citet{cappellari2002}.

In this study, we use the MGE parametrisation of the photometry to construct our dynamical models. These models assume axisymmetry, but galaxies can show non-axisymmetric features, such as bars. \citet[Sec.~3.2.1]{Scott2013} tried to reduce the effect of bars in the models by essentially ignoring them during the fits, while forcing the MGE to only describe the light distribution of the underlying galaxy disk. In this study, we also adopted the same approach for all galaxies.

The luminosity of the galaxies are calculated from the analytic sum of the Gaussian luminosities equation~(10) of C13b, while the projected half-light radii (\re) were computed from the circularised Gaussians using equation~(11), of C13b. For our galaxies, we also calculate the $R_{\rm e}^{\rm Maj}$, the major axis of the isophote containing half of the light of the galaxy as these are found to be a more robust measurement of the galaxy size due to its lower dependency on galaxy inclination \citep[][C13b]{Hopkins2010}. We derive this value as described in section~3.3.1 of C13b. 

In \citet{ShettyCappellari2014ApJL} we show that, for the subset of 68 galaxies for which we construct dynamical models, the \re\ derived using the MGE parametrization are consistently offset from the \re\ one can predict from the virial equation $M_{\rm vir}=5.0\re\sigma_{\rm e}^2/G$, when using the coefficient of \citet{cappellari2006}. We found that those galaxies required a 0.16 dex increase in \re, for $M_{\rm vir}$ to match the $M_{\rm JAM}$ derived via JAM models. Since we use the same data, at the same redshift, for all 154 galaxies, we apply the same fractional correction to the \re\ and $R_{\rm e}^{\rm Maj}$ of all galaxies of this study. Since the F814W filter approximates to the B band at the redshift range of our galaxies, we state all luminosity and photometric quantities in the B band.

\subsection{Dynamical Modelling}
\label{sec:Dyn}

To further understand the evolution of galaxies since $z\sim1$, we created dynamical models for a subsample of our galaxies. In this study we use the Jeans Anisotropic Modelling (JAM; see footnote 1) method and code of \citet{jam}, which solves the \citet{Jeans1922} equations under the assumptions of axisymmetry, while allowing for orbital anisotropy and rigorously accounting for seeing and aperture effects.

The models require as input an accurate description of the projected light distribution of the galaxy and a parametrization for the unknown total mass. As the galaxy inclination is generally unknown, we assume a standard inclination of 60$^{\circ}$, the average inclination for a set of random orientations. When this inclination is not allowed by the MGE model, we adopt the lowest allowed one. Here an inclination of 90$^{\circ}$  corresponds to an edge-on view. Another quantity required by JAM is the velocity anisotropy $\beta_{\rm z}\equiv1-\sigma_z^2/\sigma_R^2$, where $\sigma_z$ and $\sigma_R$ are the velocity dispersion along the symmetry axis and along the cylindrical radius, respectively. In this study, we adopt $\beta_z=0.2$ as this is found to be a typical value for ETGs in the local universe \citep{Gerhard2001,sauron10}. Tests by \citet{cappellari2006} and \citet{lablanche2012} show that the derived M/L through the JAM code is weakly sensitive to the effects of inclination and spatially constant velocity anisotropy.

The 68 galaxies in our modelling subsample are the same for which we created dynamical models in \citet{ShettyCappellari2014ApJL}, and we adopt the stellar velocity dispersion from that study. The stellar kinematics were derived from the DEEP2 spectra within a 1 arcsec slit. In \citet{ShettyCappellari2014ApJL} our JAM models assumed the total mass is distributed like the luminous one, hence our derived dynamical M/L incorporates the effect of stellar and dark matter.
Here we improve our dynamical models by explicitly including dark matter using results from the abundance matching techniques. This infers a relation between the stellar and halo mass by matching simulated dark halo mass functions to observed galaxy luminosity functions, under the assumption that the most massive galaxies reside in the most massive halos \citep[e.g.][]{Kravtsovetal2004,Conroyetal2006,ValeOstriker2006,ValeOstriker2007}.

Here we try two different relations between the stellar and halo mass. The first is taken from \citet{Leauthaudetal2012}, where the authors fit a log-normal Stellar-to-Halo mass relation (SHMR), with a log-normal scatter, to the results of weak galaxy-galaxy lensing, galaxy clustering and galaxy number density analysis of the COSMOS survey \citep{COSMOS}. We use equation~(13) of \citet{Leauthaudetal2012} and the parameter values for model \textit{SIG\_MOD1} within the relevant redshift bin to derive the halo mass associated with a galaxy of given stellar mass. The second SHMR we use is equation~(2) of \citet{MosterNaabWhite2013}. Here the authors use high resolution N-body simulations to derive the dark halo mass function at different redshifts, and match them to the observed galaxy stellar mass functions, hence allowing one to derive the halo mass for a galaxy of a given stellar mass.

We further assume that the dark matter of our galaxies is distributed spherically with an NFW profile \citep{NFW_DarkHalo_1996}. We constrain the concentration of the halo using equation~(12) of \citet{KlypinTrujillo-GomezPrimack2011}, where the authors study the evolution of the concentration of the galaxies with redshift using the Bolshoi simulation. Using the above equations, we calculate the NFW density profile, which is fit using the one dimensional MGE fitting code of \citet[see footnote~1]{cappellari2002}, to use in JAM.

In practice, starting from a trial $(M_{\ast}/L)_{\rm JAM}$ for a galaxy, we compute the total stellar mass by multiplication with the total luminosity. We then associate a corresponding dark halo mass using the SHMR, and a NFW profile using the halo mass versus concentration relation mentioned above. For given stellar and dark matter distributions we use JAM to predict the stellar $V_{\rm rms}=\sqrt{V^{2} + \sigma^{2}}$, integrated within the adopted slit and convolved by the seeing. The model $V_{\rm rms}$ is matched to the observed velocity dispersion by varying the only free parameter, $(M_{\ast}/L)$ of our model, using a root finding algorithm. In this manner we account for the dark matter content of the 68 galaxies in our sample, and derive their dynamical stellar $(M/L)$ ($(M_{\ast}/L)_{\rm JAM}$).

\section{Results}

\subsection{Observed Star Formation History}

\begin{figure}
  \centering
	\includegraphics[clip, width=0.95\linewidth]{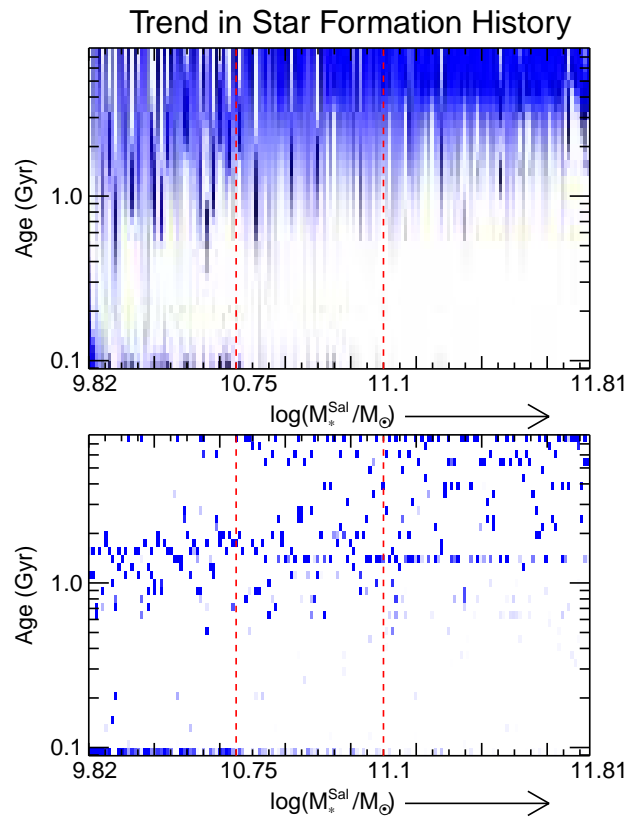}
	\caption{These plots illustrate the fraction of mass represented by stellar populations of different ages for each galaxy. Along the y-axis are stellar populations of different ages, while along the x-axis are the stellar masses of the galaxies, sorted such that the stellar masses increases from the left to right. The intensity of the blue shade represents the fraction of mass of the galaxy composed of stellar populations of a certain age. The top panel represents the SFH of the galaxies derived through the mass-weighted regularised full-spectrum fitting of the galaxy spectra to the entire template grid, while the bottom panel represents the mass distribution from a mass-weighted unregularised full spectrum fitting to the stellar populations with only solar metallicities. Both plots demonstrate that galaxies with lower stellar mass tend to have a significant fraction of young stellar populations unlike those with the highest stellar masses where the bulk of their stellar populations are formed very early on in the history of the universe. This correlation is further demonstrated in Fig.~\ref{Weight_coadd}, where we have coadded the galaxy spectra based on their stellar mass. The bin sizes of these stellar mass bins are represented here in vertical red dashed lines at stellar masses.}
	\label{SFH}
\end{figure}

In Fig.\ref{SFH}, we have plotted the SFH of the galaxies after having collapsed them along the metallicity axis. The galaxies are sorted along the x-axis of the plot in order of increasing stellar mass as derived by stellar population modelling. In the top panel, we present the regularised SFH and we can clearly see that there is a gradual trend in the presented SFH with stellar mass. It appears that galaxies with low stellar mass tend to have a significant mass fraction of their stellar populations younger than 1 Gyr, unlike the typical SFH of high stellar mass galaxies, where the significance of this young stellar population weakens. 

For comparison in the bottom panel we have presented the unregularised mass distribution for the full-spectrum fitting of the galaxies. In this case the weights distribution is much noisier and consists of discrete peaks, as expected. But the main trend in SFH with galaxy mass can still be recognized. This shows that the recovered trend is robust.

\begin{figure*}
  \centering
	\includegraphics[clip, width=0.9\linewidth]{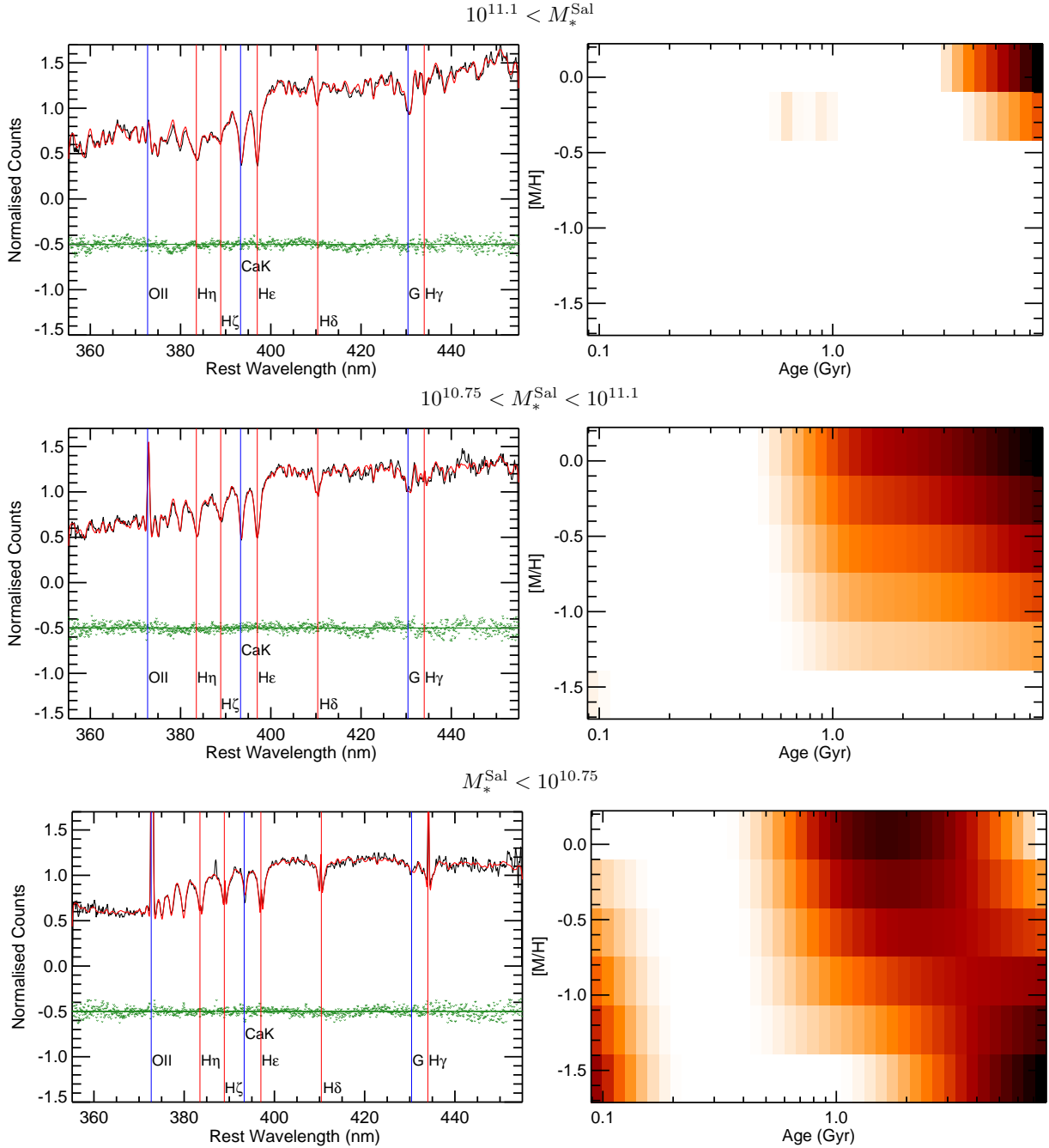}
	\caption{Here we present the coadded spectra for the three stellar mass bins: $10^{11.1} < M^{\rm Sal}_{\ast}$, $10^{10.75} < M^{\rm Sal}_{\ast} < 10^{11.1}$ and $M^{\rm Sal}_{\ast} < 10^{10.75}$. The plots on the left hand side illustrate each coadded spectra (Black line) along with the best fit from pPXF (Red line) plotted on top. The green points below are the residuals of the fit vertically shifted to -0.5. For reference, we have also plotted the location of stellar absorption and gas emission lines within the observed wavelength. By eye it is clear that the quality of the coadded spectrum is far superior to that observed for each individual spectrum. In this case, the pPXF fit includes both stellar SSP models and gas emission line templates. For the latter, we included in the fits the Balmer emissions and the [OII]$\lambda$3727 doublet, for which we adopted the same kinematics. This allows us to use the full spectral range without the need for masking the emission lines. Along the right hand side, we have plotted the mass fraction of the fitted stellar population models of various ages and metallicities. We clearly see a trend between the star formation history of the galaxies along with the stellar mass of the galaxies. The coadded spectra of the highest masses indicate that these galaxies are likely to form the bulk of their stars in the early epochs of the universe and undergo a passive evolution thereafter. In the case of the intermediate masses, the galaxies appear to have an extended SFH while the lowest mass spectra require recent star formation, along with a very old population, to match the observed spectra.}
	\label{Weight_coadd}
\end{figure*}

To further illustrate this trend, we have coadded the galaxy spectra into 3 bins: $10^{11.1} < M^{\rm Sal}_{\ast}$, $10^{10.75} < M^{\rm Sal}_{\ast} < 10^{11.1}$ and $M^{\rm Sal}_{\ast} < 10^{10.75}$. The full-spectrum fitting of these coadded spectra, along with their associated SFH, are shown in Fig.~\ref{Weight_coadd}. As the S/N is much higher for the coadded spectrum of the galaxies, we adopt the standard $\Delta\chi^2=\sqrt{2\times N_{\rm DoF}}$ criterion for regularization (for further information, please refer to Section~\ref{sec_SFH}). 

Also, the significantly high S/N allows us to confidently fit the coadded spectra using the entire metallicity range of the MILES models, [M/H] = -1.7 to 0.22. We should point out that the [M/H] is not equivalent to [Fe/H] for the low metallicity models, however \citet{milesmodels} state that the this doesn't effect the galaxy age and metallicity estimates significantly. The same trend that we could see in the individual spectra appears cleaner in the stacked ones: (i) the most massive galaxies require a maximally old and nearly solar population to reproduce the observations, with little room for recent star formation. (ii) The galaxies with intermediate masses still peak at the oldest ages, but allow for a more extended star formation history. (iii) The lowest mass galaxies do not peak at the oldest ages any more and additionally show clear evidence for ongoing star formation.

The well known age-metallicity degeneracy of stellar population models implies that the effect on a galaxy spectrum due to a decrease in its mean age can be negated by an increase of its metallicity \citep{Worthey1994,Renzini2006}. This degeneracy is reduced by the use of full-spectrum fitting, which tries to reproduce a large number of spectral features \citep{Conroy2013}, but the general trend is still present in our recovered age and metallicity distributions. An example of this can be seen in the bottom right panel of Fig.~\ref{Weight_coadd}: The ridge of equally-large weights going from an age of 2 Gyr with nearly solar metallicity, to an age of 7 Gyr with the lowest metallicity is likely due to the degeneracy, combined to our relatively narrow wavelength range. In Table.~1, we present the relevant stellar modelling quantities of the full galaxy sample of the study. The complete version of this table is available online with the published article.

In this study, we calculate the mean age of the galaxies from the SFH as follows:
\begin{equation}
	\langle\log({\rm Age})\rangle = \frac{\sum_{j}w_{j}\log({\rm Age}_{j})}{\sum_{j} w_{j}},
\label{age_eq}
\end{equation}
where $w_j$ is the weight associated to the $j$th template during the regularised luminosity-weighted full-spectrum fit and ${\rm Age}_j$ is the corresponding age. To obtain luminosity-weighted quantities one can simply normalize all the SSPs to the same mean flux within the fitted wavelength range, before the pPXF fit. Of course the luminosity weighting refer to the fitted wavelength, which in our case lies between the $U$ and $B$ band. When the SSP spectra used in the pPXF fit have the flux as given by the models, namely corresponding to a unitary stellar mass, this equation provides mass-weighted quantities. 

\begin{table*}
	\begin{center}
		\begin{threeparttable}
			\caption{Results of Stellar Population Modelling. The table below is a partial representation of the complete set, which is available as part of the online material of the paper on \mnras.}
			\begin{tabular}{cccccccccc}
				\hline\hline
				{DEEP2 ID} & {Redshift} & {$R_{\rm e}$} & {$R_{\rm e}^{\rm Maj}$} & {$M_{\rm B}$} & {$(M_{\ast}/L)_{\rm B}^{\rm Sal}$} & {Mass-} & {Luminosity-} & {$log(M_{\ast}^{\rm Sal})$} \\ 
				{} & {} & {} & {} & {} & {(Regul.)} & {weighted Age} & {weighted Age} & {} \\ 
				{} & {} & {(kpc)} & {(kpc)} & {(Mag.)} & {(\msun/\lsun)} & {(Gyrs)} & {(Gyrs)} & {(\msun)} \\
				{(1)} & {(2)} & {(3)} & {(4)} & {(5)} & {(6)} & {(7)} & {(8)} & {(9)} \\
				\hline
				11044628 & 0.824 & 9.7 & 10.0 & -22.06 & 0.5 & 1.65 & 0.28 & 10.686 \\
				11050845 & 0.840 & 4.4 & 5.2 & -21.76 & 3.5 & 4.85 & 4.57 & 11.429 \\
				11051246 & 0.777 & 4.4 & 4.6 & -21.18 & 0.9 & 2.47 & 0.71 & 10.592 \\
				12003915 & 0.749 & 3.2 & 5.1 & -20.89 & 0.7 & 2.52 & 0.34 & 10.391 \\
				12004067 & 0.748 & 9.4 & 10.2 & -21.76 & 0.8 & 2.26 & 0.54 & 10.784 \\
				\hline
			\end{tabular}
			\begin{tablenotes}
				\small
				\item Column (1) : DEEP2 galaxy identifier. Column (2) : DEEP2 estimated redshift. Column (3) : Effective radii of the galaxies derived analytically from the circularised MGE model. This \re ~has been corrected for the underestimation that was shown in \citet{ShettyCappellari2014ApJL}. Column (4) : Major axis of the isophote containing half of the observed light of the galaxy. Similar to the \re, this value has also been corrected for the underestimation. Column (5) : Absolute B (Vega) band magnitude of the galaxies derived analytically through MGE models of the F814W photometry. Column (6) : The mass-weighted $M_{\ast}/L_{\rm B}$ derived through the regularised full-spectrum fit using the entire template grid. In this study, we adopt the intrinsic scatter between the $(M_{\ast}/L)$ derived through the different methods and template grids as their error, i.e. 0.08 dex. Column (7) : Mass-weighted age of the galaxies calculated from the regularised full-spectra fit using the entire template grid. Column (8) : Luminosity-weighted age of the galaxies calculated from a regularised full-spectra fit using the entire template grid. Column (9): The log of the stellar mass, in units of \msun, derived using the values in Column (7) and Column (5). Adopting the intrinsic scatter between our value and that of \citet{bundy2006}, 0.16 dex is a good approximation to the error in this value.
			\end{tablenotes}
		\end{threeparttable}
	\end{center}
\end{table*}

\begin{figure}
  \centering
	\includegraphics[clip, width=0.9\linewidth]{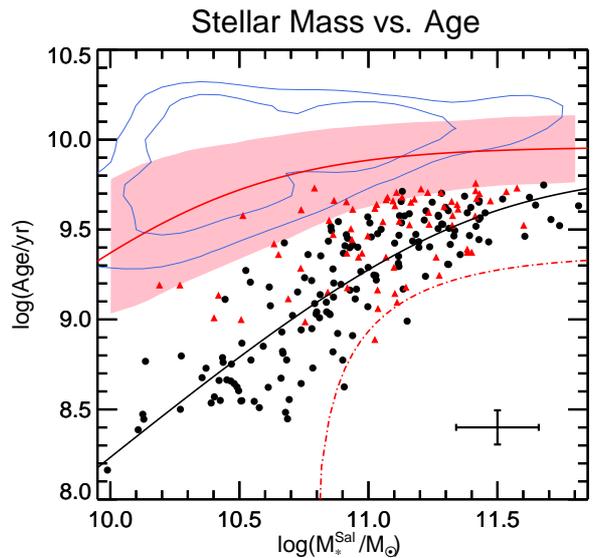}
	\caption{In this plot, we present the observed trend between the regularised luminosity-weighted stellar ages of galaxies and their stellar masses. The black points represent our sample of 154 galaxies at redshift $0.7<z<0.9$ with the solid black line representing the best fit to the trend using Eq.~\ref{age_Para}. A representative error bar for the galaxies is presented in the bottom right of the plot. The red triangles in the plot are the luminosity-weighted ages of \citet{Gallazzietal2014}. The plot demonstrates the trend between our luminosity-weighted ages and the stellar mass of the galaxies. In this plot we also present two samples of the local universe for reference. The solid red line represents the best fitting relations for the galaxies of \citet{gallazzi2006} with the red band enclosing the 16$^{\rm th}$ and 84$^{\rm th}$ percentiles of the distribution. The blue contours represent the probability density of the \atl sample. The dash-dotted red line represents the trend expected from the local galaxies of \citet{gallazzi2006} at z~$\sim0.8$ if these galaxy have evolved passively since then.}
	\label{AgePlot}
\end{figure}

In Fig.~\ref{AgePlot}, we present the luminosity-weighted ages of the galaxies against their respective stellar masses. The black points represent the full sample of 154 galaxies of this study, while the red triangles represent the luminosity-weighted ages of the entire sample of $\sim70$ galaxies of G14, at similar redshift. On the bottom right of the figure, we plot a representative error bar for the 154 galaxies in this study. The error on the stellar mass is based on the scatter in Fig.~\ref{Bundy}, ie 0.16 dex. To estimate the error on the ages we bootstrap the galaxy spectra to derive their age multiple times. From this distribution of ages for each galaxy, we calculate robust standard deviations and take the median of these as the representative error. This bootstrapping is done using only the templates with solar metallicities and without the regularisation of the template weights.

The two sample appear to largely overlap, however our sample includes a larger number of low-mass and correspondingly younger galaxies. The solid black line is the resulting best-fit to our galaxy sample using a similar formula as G14:
\begin{equation}
	\frac{\rm Age_b}{{\rm Age}(M_\ast)} = 1+\left(\frac{M_b}{M_\ast}\right)^\gamma,
\label{age_Para}
\end{equation}
which describes an age trend as a function of stellar mass $M_\ast$ following a power-law ${\rm Age}\propto M^{\gamma}_\ast$ at small masses and reaching the asymptotically constant value ${\rm Age_b}$ above the characteristic mass $M_b$, with a smooth transition in between. The best-fitting coefficients for the luminosity-weighted and mass-weighted (not shown) ages of our 154 galaxies are presented in Table.~2.

In Fig.~\ref{AgePlot}, we also present two local samples. The median luminosity-weighted age trend of the sample of both spirals and ETGs from \citet{Gallazzietal2005} is represented by the red solid line, with the red band enclosing the $16^{\rm th}$ and $84^{\rm th}$ percentiles of the age distribution. The blue contour lines represents the probability density distribution of the single stellar population ages (similar to luminosity-weighted ages) of the volume-limited 260 ETGs of the \atl\ survey \citep{atlas3d1}, as determined by \citet[Hereafter MD15]{ATLAS3D30}. This was calculated using the Epanechnikov kernel, with a smoothing template that was optimised for each dimension, based on \citet{silverman1986}. The dot-dashed red line in the plot represents for reference the passive back evolution of the trend observed in \citet{Gallazzietal2005}. These two samples provide a local universe reference for our results, and given the differences in the two samples selection give a sense of the possible ranges in the local trends. Our most massive galaxies are {\em not} consistent with passively evolving into the corresponding sample of spirals and ETGs of the \citet{Gallazzietal2005} sample, but they {\em are} consistent with evolving passively into the massive ETGs of MD15. This is not only true according to their mean ages but, most importantly, it is also true also for the actual SFH, which in both cases shows the galaxies essentially consists of a maximally old stellar population.

The results presented in Fig.\ref{SFH} and Fig.\ref{Weight_coadd} are in line with the results of G14, where the authors predict that the SFH of galaxies has to be a function of their stellar mass. We note that since our analysis is based on the spectra of galaxies taken within the central ~1$\arcsec$ of the galaxies, these results are relevant only to that region. Our analysis is unable to give any direct insight into the evolution of the outer regions of these galaxies.

\begin{table}
	\begin{center}
		\begin{threeparttable}
			\caption{Best fitting parameters for equation~\ref{age_Para}}
			\begin{tabular}{cccc}
				\hline\hline
				{Technique} & {$\log{\rm Age}_b$} & {$\log{M_b}/\msun$} & {$\gamma$}\\
				\hline
				Mass-weighted Ages & 9.70 & 10.66 & 1.15 \\
				Luminosity-weighted Ages & 9.83 & 11.35 & 1.18 \\
				\hline
			\end{tabular}
		\end{threeparttable}
	\end{center}
\end{table}

By directly studying the SFH of galaxies at half the age of the universe, our study serves as a stringent test to the scenario of galaxy formation and evolution that was presented in the Introduction of this paper. Work by \citet{Onoderaetal2012}, where the authors use stacked spectrum of 17 passively evolving galaxies at redshifts of $z\gtrsim1.4$, demonstrated that those massive galaxies had been quenched since $z\sim2.5$. This is consistent with work by \citet{Juneauetal2005} and \citet{Behroozietal2013}, where the authors use observational data covering a large redshift range, to show that the star formation rate peaks at redshift of $\sim3$ for the most massive galaxies and that this peak evolves to lower redshifts for lower mass galaxies. 

This trend in the star formation rate of the galaxies has been known for about a decade from observations of the fossil record in the local Universe \citep[e.g.][MD15]{Heavensetal2004,Thomas_etal_2005,Panteretal2007}. However the local studies require a significant extrapolation to describe the SFH over the full evolution of the Universe. While our study directly recovers the SFH at an epoch when the Universe was half its current age, it is reassuring to find that the extrapolated local SFH trends are consistent with what we observe. Our result further strengthens these results and the picture of galaxy evolution that they support, by demonstrating that the bulk of stars of the massive galaxies had formed in the early epochs of the Universe and have been evolving passively, while low mass galaxies were/are still forming significant mass fractions of their stellar populations till $z\sim0.8$. 

\begin{figure*}
  \centering
	\includegraphics[clip, width=0.9\linewidth]{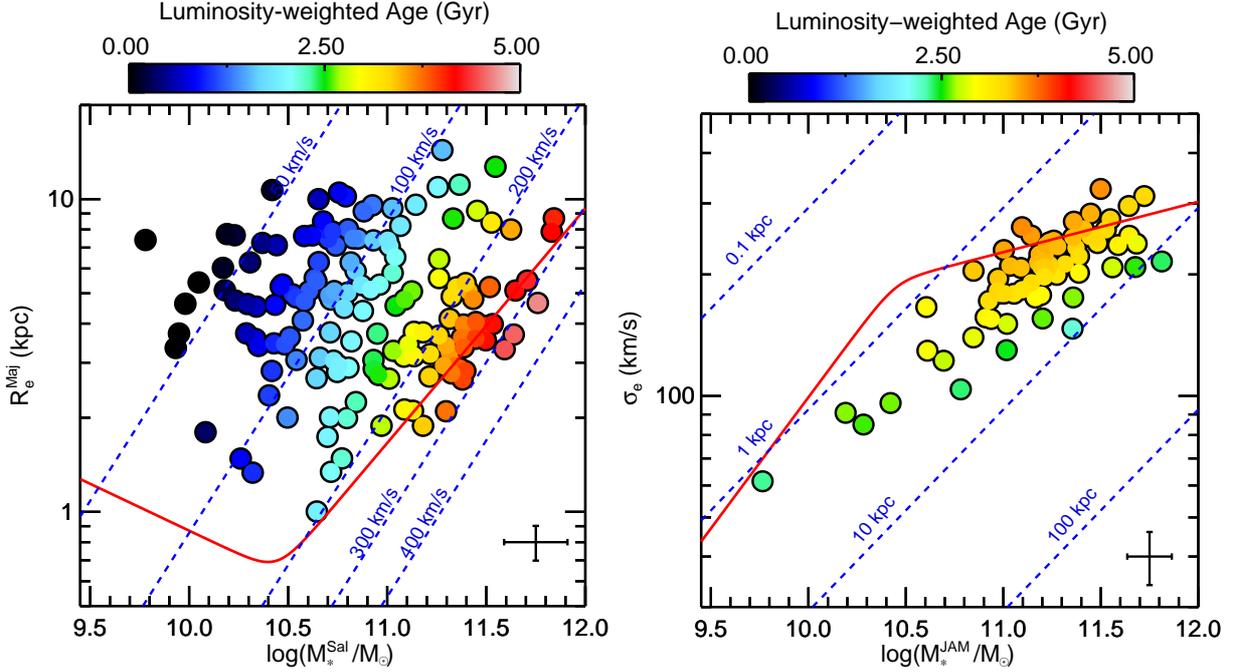}
	\caption{Here we present in the left panel the mass size plot for the 154 galaxies in our full galaxy sample. The mass in the plot is derived from the mass weighted regularised stellar population modelling while the galaxy size is represented by $R_{\rm e}^{\rm Maj}$. The solid red line of the plot defines the Zone of Exclusion (ZOE) determined by the \atl group (C13a), beyond which the probability distribution of galaxies is negligible. The dashed blue lines are lines of constant velocity dispersion. The colour trend observed in the plot is the LOESS smoothed (C13b) luminosity weighted age. This plot looks similar to its dynamical variant in the local universe (See fig.~6 and fig.~7 of MD15), where the authors find that the age of the stellar population within a galaxy is correlated to the its velocity dispersion. The left panel also demonstrates that some quiescent galaxies are on the other side of the ZOE and would require to grow by a factor of $\sim1.5$ to be consistent with local observations. In the right hand panel, we present the dynamical stellar mass, derived using the $(M_{\ast}/L)_{\rm JAM}$, against the velocity dispersion for our secondary galaxy sample, which is equivalent to the Faber-Jackson relationship. In this plot, the dashed blue lines depict lines of constant size. Since the quantities of this plot is derived in a manner independent of those in the left panel, this plot illustrates the robustness of the results discussed. In the lower right corner of both panels, we present a representative error bar for the quantities.}
	\label{MP}
\end{figure*}

\subsection{Mass-Size Distribution}

In C13b, the authors demonstrate that by replacing the luminosity with mass, and the circularised \re~with $R_{\rm e}^{\rm maj}$, the Fundamental Plane evolves into a Mass Plane, which follows the virial equation. This shows that the Fundamental Plane is due to virial equilibrium, combined with a smooth $M/L$ variation with the galaxy velocity dispersion $\sigma$. Most of the information on galaxy evolution is contained in the non edge-on views of the plane, e.g.\ the Mass-size and Mass-$\sigma$ distribution. \citet[Hereafter C13a]{atlas3d20} shows that galaxy properties, namely the stellar age, colour, $M/L$, IMF and gas content, all best follow lines of constant $\sigma$, which they demonstrate traces the bulge mass fraction. MD15 additionally showed that metallicity and non-solar abundance are also driven by $\sigma$. Below we present the trend of the stellar population in the mass plane of galaxies at $z\sim1$ and compare it with that observed in the local universe.

Using the physical parameters derived in this study we are able to study the mass plane of our galaxies which are presented in Fig.\ref{MP}. In the left panel of Fig.~\ref{MP}, we plot the stellar mass, derived through stellar population modelling, and the galaxy size for the 154 galaxies in our full sample. In the lower right corner, we plot a representative error bar for the derived quantities. The error on the stellar mass derived using stellar populations modelling is based on the scatter observed in Fig.~\ref{Bundy}, ie 0.16 dex. The error on the galaxy size is based on our comparisons with \citet{fernandezlorenzo2011}, where the authors fit a de Vaucoleurs stellar profile to I-band photometry of 40 galaxies present in our sample. We compare the effective radii calculated by these authors with that calculated using the MGE models of this study and derive an error of 0.05 dex on the galaxy size. The colour of the symbols in the two plots represents the luminosity-weighted age (Gyrs) of the stellar population in the galaxy. This was LOESS smoothed \citep{cleveland1979} using implementation (see footnote 1) of C13b. This plot can be directly compared to fig.~1 and fig.~3 of C13a and to fig.~7 of MD15, where the authors respectively show that for the \atl\ volume limited sample of ETGs the observed stellar population indicators and stellar age of the galaxies correlate with their velocity dispersion. Our plot suggests that this trend is in place by $z\sim1$. 

The solid red line in the left panel of Fig.~\ref{MP} is the Zone of Exclusion (ZOE) for the nearby galaxies as prescribed by C13a, beyond which the probability of finding galaxies within a volume-limited sample is $\sim1\%$. The left panel demonstrates that some galaxies in the full sample would lie a factor of $\sim1.5$ below the ZOE in the local universe. The robustness of this result is demonstrated in the right panel of Fig.~\ref{MP}, where we have plotted the dynamical stellar mass against the stellar velocity dispersion for the secondary galaxy sample. This plot is equivalent to the Faber-Jackson relationship with the galaxy luminosity substituted for the dynamical stellar mass derived by the multiplication of $(M_{\ast}/L)_{\rm JAM}$ with the total luminosity of the galaxies. The solid red line in this plot is the ZOE translated into this Faber-Jackson plot via the Virial equation. Despite the stark contrast between the methodology used to derive the physical parameters in the two plots, both plots present a consistent picture. Considering that the \atl\ sample is volume-limited and thus provides a representative reference for the local population, while our $z\sim1$ sample is biased towards the most dense objects (Sec.~\ref{bias}), this implies that the factor of 1.5 provides a stringent upper limit to the to the growth of galaxies since $z\sim1$. To aid the reader in accurately interpreting this plot, a representative error is presented. This error bar is based on the median error of the velocity dispersion, derived via bootstrapping (see section 3.1 of \citet{ShettyCappellari2014ApJL}), and the propagation of this error onto the derived dynamical mass via eq.~8 of \citet{jam}.

Our upper limit to the size growth of the most compact galaxies since $z\sim1$ is consistent with previous works such as \citet{Newmanetal2012}, where the authors fit the observed mass-size relation for galaxies in bins of redshift. Their fits suggest that a galaxy of mass $M_{\ast}=10^{11}\msun$ within their redshift bin of 0.4 to 1 grow by a factor of 1.2 by redshift of 0.06. In the work presented by \citet{vanderWeletal2014}, the authors use 3D-HST and CANDELS data to study the evolution of the mass-size distribution of 30,958 galaxies with stellar masses $>10^{9}\msun$ and $0<z<3$. Their comparison of the size evolution of the median sizes of galaxies from $z=0.75$ to $z\sim0$ demonstrates a size growth of $\sim1.4$, consistent with our findings. There have been recent suggestions that part of the observed size growth in quiescent galaxies may be due a progenitor bias, i.e. the increase in the average size of the quiescent galaxy population with cosmic time is due to the inclusion of larger recently quenched galaxies into the quiescent galaxy population \citep[e.g.][]{Poggiantietal2013,Carolloetal2013}. However, the results of recent work \citep[e.g.][]{Krogageretal2014,Bellietal2015} have shown that the progenitor bias isn't sufficient to accommodate the size growth that has been observed since $z\sim2.5$.

\subsection{Dark Matter Fraction and IMF}

\begin{figure}
  \centering
	\includegraphics[clip, width=0.9\linewidth]{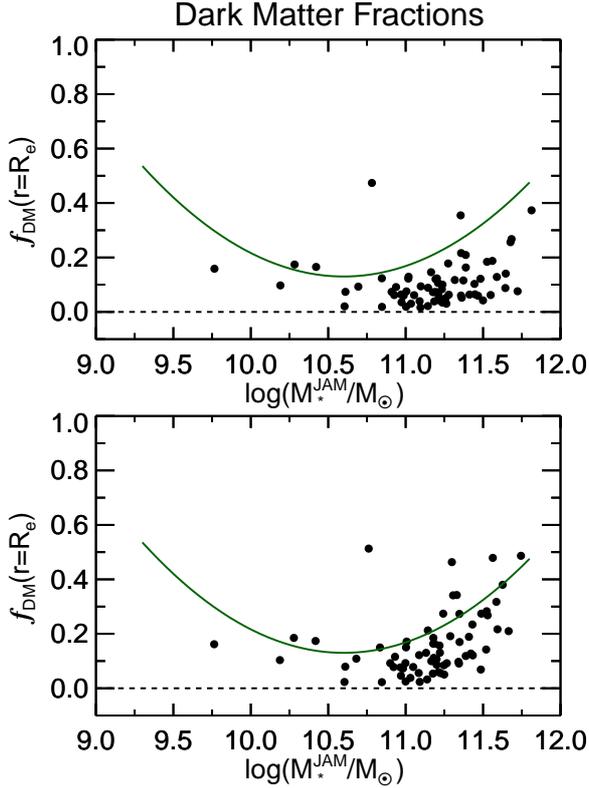}
	\caption{Here we present the dark matter fraction within 1\re~for the 68 galaxies for whom we have created dynamical models. The dark matter fraction is the dark matter enclosed within a radius of 1\re~divided by the total mass (i.e. stellar + dark matter) enclosed by that radius. In the top panel, the dark matter haloes are assigned based on the SMHR derived by \citet{Leauthaudetal2012} while the SMHR by Moster et al (2013) is used to derive the dark matter haloes for the bottom panel. Despite the difference in the absolute value of the dark matter fraction, both results demonstrate a low dark matter fraction for the inner regions of massive galaxies.}
	\label{DM_frac}
\end{figure}

\begin{figure}
  \centering
	\includegraphics[clip, width=0.9\linewidth]{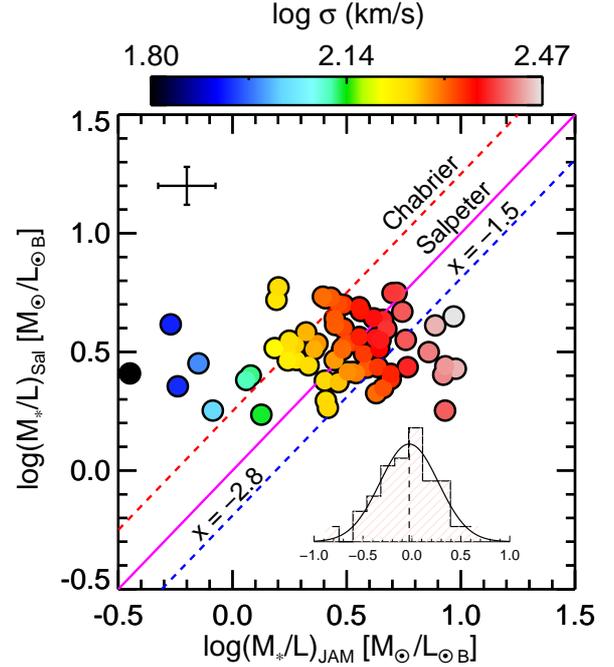}
	\caption{Here we present the IMF normalisation for the 68 galaxies with dynamical models. This is an updated version of the fig.~4 in \citet{ShettyCappellari2014ApJL} the difference being that here we use the $(M_{\ast}/L)_{\rm JAM}$ derived from dynamical models, instead of the total one. Along the x-axis, we have plotted the $(M_{\ast}/L)$ derived via dynamical modelling while the y-axis represents the stellar $(M_{\ast}/L)$ derived via stellar population modelling, assuming a Salpeter IMF. The colour trend shown here is the LOESS smoothed (C13b) central velocity dispersion of the galaxies. Representative errors on the $(M_{\ast}/L)$ is presented in the top left corner of the plot, based on Fig.~\ref{Bundy} and eq.~8 of \citet{jam}.}
	\label{IMF_norm}
\end{figure}

In \citet{ShettyCappellari2014ApJL}, we had modelled the stellar kinematics of the galaxies assuming that mass-follows-light and using the total $M/L$ to account for the dark matter contribution. When making our inferences on the IMF of those galaxies we made the empirically-motivated assumption that the dark matter fractions within their central regions were negligible. The basis for our assumption were mainly the results of C13b and \citet{Hilz2013}. In C13b, the authors were able to derive the dark matter fraction of their 260 local ETGs using integral-field data and dynamical modelling, and accounting for the dark matter via multiple dark matter profiles. Their results robustly proved that the dark matter fraction of the central regions of massive ETGs is \textless 30 per cent. Furthermore, galaxy merger simulations of \citet{Hilz2013} find that the dark matter fraction of galaxies would increase with time, i.e. with decreasing redshift. 

A robust test of the assumption would be a direct measurement of the dark matter fractions in these galaxies, however this is not feasible from the stellar dynamics without spatially-resolved information, which is still difficult to obtain with current instruments. Therefore, to test this assumption, we place our galaxies within dark matter haloes predicted by abundance matching techniques. Furthermore, to get robust results on the dark matter fractions of the galaxies, we have derived the dark matter fractions for our galaxies using two independent SHMRs (Sec.~\ref{sec:Dyn}).

The results of the dynamical modelling are presented in Table.~3 and illustrated by Fig.~\ref{DM_frac}. The plots demonstrate that, to be consistent with the observed surface brightness and velocity dispersion, as well as with the halo matching results, massive galaxies at this redshift must contain low dark matter fraction $f_{\rm DM}(\re)$ within a sphere of radius 1\re. The inferred median value for our sample of 68 galaxies is $f_{\rm DM}(\re)=9$ and 13 per cent, while the 90th percentile is $f_{\rm DM}(\re)=21$ and 34 per cent, using the SHMR of \citet{Leauthaudetal2012} and \citet{MosterNaabWhite2013} respectively. 

In the two panels of Fig.~\ref{DM_frac}, the green line represents the robust parabolic fit to the dark matter fractions of the 260 galaxies (C13b) derived using SHMR relation of \citet{Moster_etal2010}, while additionally requiring the models to best fit the integral-field kinematics and photometry. This is similar to our dark matter fraction results derived using \citet{MosterNaabWhite2013}. Comparing our results to that of \atl, the dark matter fraction appears to have increased in the last $\sim7$~Gyrs. The lower dark matter fraction derived at higher redshifts is consistent with the numerical simulation prediction of \citet{Hilz2013} and is also consistent with recent results by \citet{Beifiorietal2014}, which found that the ratio of the dynamical mass (dark matter + stellar mass) to the stellar mass of the galaxies reduces with increasing redshift. However, we do note that this trend may be in part influenced by the differences in the sample selection.

\begin{table*}
	\begin{center}
		\begin{threeparttable}
			\caption{Results of Dynamical Modelling. The table below is a partial representation of the complete set, which is available as part of the online material of the paper on \mnras.}
			\begin{tabular}{cccccc}
				\hline\hline
				DEEP2 ID & $\sigma$ & $\delta\sigma$ & $(M_{\ast}/L)_{JAM}$ & $log(M_{\ast}^{\rm JAM})$ & \textit{f}$_{DM}$ \\
				{} & {(\kms)} & {(\kms)} & {(\msun/\lsun)} & {(\msun)} & {} \\
				{(1)} & {(2)} & {(3)} & {(4)} & {(5)} & {(6)} \\
				\hline
				11050845 & 298.2 & 28.7 & 5.74 & 11.644 & 0.088 \\
				12004136 & 202.4 & 21.7 & 2.80 & 11.275 & 0.178 \\
				12004516 & 176.4 & 28.6 & 2.56 & 11.164 & 0.146 \\
				12008254 & 250.7 & 33.3 & 4.56 & 11.444 & 0.103 \\
				12008360 & 185.2 & 31.1 & 2.54 & 10.974 & 0.035 \\
				12008441 & 257.7 & 64.3 & 8.53 & 11.646 & 0.141 \\
				12011900 & 318.5 & 57.1 & 9.30 & 11.499 & 0.042 \\
				\hline
			\end{tabular}
			\begin{tablenotes}
				\small
				\item Column (1) : DEEP2 galaxy identifier. Column (2) : Velocity dispersion as measured by full-spectrum fitting of the galaxy spectra. Column (3) : Error on the derived velocity dispersion of the galaxy through a bootstrapping technique. Column (4) : B-band $(M_{\ast}/L)$ derived through dynamical modelling, where the dark matter halo was assigned using the SHMR of \citet{Leauthaudetal2012}. Column (5) : The log of the stellar mass, in units of \msun, of the galaxy derived using the absolute luminosity (Column (5) of Table.~1) and the M/L of Column (4). Column (6) : Fraction of dark matter within 1\re ~of the galaxy. This was derived by using the SHMR of \citet{Leauthaudetal2012}.
			\end{tablenotes}
		\end{threeparttable}
	\end{center}
\end{table*}

In the work presented in \citet{ShettyCappellari2014ApJL}, the IMF normalisation of the 68 dynamically modelled galaxies was shown in fig.~4. The plot illustrated that the IMF normalisation for massive ETGs at $z\sim1$ was Salpeter-like, however a caveat of the work was the assumption that the dark matter fractions in the central regions of high redshift galaxies was not significant. In this study we include dark matter explicitly and in Fig.~\ref{IMF_norm} present an updated version of the plot, with the dark matter contribution explicitly removed. This confirms the assumption that dark matter fraction has a negligible effect on the IMF normalisation of the galaxies, i.e. the {\em average} IMF normalisation of massive ETGs at redshift of $\sim 1$ is still consistent with a Salpeter IMF. This result is independent of the SHMR used. 

This confirmation of our \citet{ShettyCappellari2014ApJL} result is also in agreement with recent work by \citet{Sonnenfeldetal2015}, where the authors have used a joint lensing and stellar dynamics modelling of massive galaxies out to $z\approx0.8$, including about ten galaxies within our resdhift range. The authors found that the stellar IMF normalisation was close to Salpeter IMF for $M_{\ast}=11.5$. Also, work by \citet{MartinNavarroetal2014}, using IMF sensitive spectral features, determined that more massive galaxies have a bottom heavier IMF at redshifts between 0.9 and 1.5. The Salpeter normalization of the IMF is also consistent with the non-universality of the IMF reported in the nearby Universe \citep[e.g.][]{vandokkum2010,cappellari2012nature}. In fact, if the centres of massive galaxies at $z\sim1$ passively evolve into the massive ETGs population, their IMF is expected to have a Salpeter normalization as observed locally \citep{auger2010b,Conroy2012,atlas3d20}.

\section{Summary}

In this study, we have determined the SFH of the galaxies at redshift $0.7 < z < 0.9$ in a non-parametric manner, using full-spectrum fitting. From a parent sample of $2,896$ galaxies from the DEEP2 survey, we apply strict quality selection criteria to extract 154 galaxies with $10^{10}\lesssim M_{\ast} \lesssim 10^{12} \msun$ for which we derive our SFHs. For a subsample of 68 galaxies, with $M_{\ast}\gtrsim 10^{11}\msun$, we additionally construct dynamical models. Due to these selection criteria, our galaxy samples have a higher magnitude cut-off than the DEEP2 survey and preferentially selects galaxies with high surface brightness. However, we show that the full galaxy sample is representative of the red sequence and all but the bluest galaxies of the blue cloud, while the secondary sample mainly consists of galaxies from the red sequence.

The derived SFH for the full galaxy sample indicates that the most massive galaxies formed the bulk of their stars in the early epoch of the universe unlike low mass galaxies which are forming stars at a significant rate even at redshift of $\sim1$. This is qualitatively consistent with previous fossil record studies in the local universe, where authors have found that the star formation rate of the most massive galaxies peaked early in the age of the universe, and hence provides a robust and independent test of these results and the narrative of the formation and evolution of galaxies that these results have produced. Our results based on Fig.~\ref{SFH} and Fig.~\ref{Weight_coadd} demonstrate that the difference between the SFH of galaxies evolves gradually and is a function of stellar mass.

We study the distribution of galaxy ages on the mass-size diagram. This demonstrates that the velocity-dispersion dependence in the age of the stellar populations of the central regions of the galaxies was already in place by $z\sim1$. We also place an upper limit of a factor of $\approx1.5$ to the size growth of individual galaxies since $z\sim1$, in agreement with other studies.

Finally, using the dynamical models of the 68 galaxies in our secondary sample, which account for the dark matter in the galaxies using results from abundance matching techniques, we measure a median dark matter fraction $f_{\rm DM}(\re)\approx10$ per cent per cent, within a sphere of radius \re, for the most massive galaxies, with small variations depending on the adopted Stellar-Halo mass relation. Comparing this to the dark matter fraction determined locally, we find that the dark matter fraction of galaxies has marginally increased in the last 8 Gyrs, but is otherwise insignificant. This result confirm the study of \citet{ShettyCappellari2014ApJL} stating that the average IMF normalisation of the most massive galaxies is on average consistent with a Salpeter IMF. 

\section*{Acknowledgements}

M.C. acknowledges support from a Royal Society University Research Fellowship.

\bibliographystyle{mn2e}

\label{lastpage}

\end{document}